\documentclass{aa}
\usepackage{txfonts}
\usepackage{graphicx}
\usepackage{txfonts}
\usepackage{natbib}
\usepackage{ulem}
\usepackage{booktabs}

\usepackage[colorlinks=true, linkcolor=blue, urlcolor=blue, citecolor=blue]{hyperref}

\usepackage{orcidlink}

\def\teff{$T_{\rm eff}$}

\def\ll_lsun{log$({L/\rm L_{\odot}})$~}  
  
\def\masa_msun{$M/ \rm M_{\odot}$~}  
  
\def\m_mstar{$M/M_{*}$~}

\begin{document}

\title{Post-common Envelope Evolution of Helium-core White Dwarfs\thanks{The cooling sequences are publicly available at \protect\url{http://evolgroup.fcaglp.unlp.edu.ar}.}}

\author{
  Leandro G. Althaus\orcidlink{0000-0003-2771-7805}\inst{1}
  \and Leila M. Calcaferro\orcidlink{0000-0002-1345-8075}\inst{1}
\and Alejandro H. Córsico\orcidlink{0000-0002-0006-9900}\inst{1}
\and Warren R. Brown\orcidlink{0000-0002-4462-2341}\inst{2}
}

\offprints{}

\institute{
Grupo de Evolución Estelar y Pulsaciones, Facultad de Ciencias Astronómicas y Geofísicas, Universidad Nacional de La Plata, CONICET-IALP, Paseo del Bosque s/n, 1900 La Plata, Argentina
\\ \email{althaus@fcaglp.unlp.edu.ar}
\and
Center for Astrophysics \textbar\ Harvard \& Smithsonian, 60 Garden Street, Cambridge, MA 02138, USA
}

\date{Received}

\abstract
{He-core white dwarfs (He WDs) from the common envelope (CE) channel 
  offer insights into binary evolution and compact remnant formation. Their 
  cooling rates influence their detectability and affect age estimates of close 
  binaries. Compared with those from stable Roche-lobe overflow (SRLOF), CE He WDs 
  experience a distinct mass-loss history, leading to fundamental differences 
  in the post-CE evolution of the resulting WDs.}
{We investigate how the H envelope mass ($M_{\rm H}$) affects
  the cooling evolution of CE He WDs. In particular, we analyze how the
  bifurcation point, which separates the degenerate He core from the
  envelope, sets the remaining $M_{\rm H}$ and the presence of residual
  H burning.}
{We computed evolutionary sequences for He WDs of $0.20 \, M_{\odot}$ to $0.42 \, M_{\odot}$, 
  from a $1 \, M_{\odot}$ progenitor on the red giant branch. Using the LPCODE stellar evolution 
  code, we followed their evolution from the post-CE phase to the cooling track, identifying two 
  pathways depending on the remaining H: (i) non-flashing sequences, where WDs cool without 
  prior nuclear burning, and (ii) flashing sequences, where H shell flashes reshape the 
  envelope before cooling.}
{CE He WDs with minimal $M_{\rm H}$ cool rapidly after formation, with 
  negligible residual H burning. For a sample with $T_{\rm eff}$ between 
  $12,000$ and $27,000$ K, our models predict ages of $5$–$130$ Myr, increasing 
  to slightly above $300$ Myr for $T_{\rm eff} < 10,000$ K, much younger than 
  those from SRLOF sequences. In contrast, WDs with more $M_{\rm H}$ sustain 
  residual nuclear burning, delaying cooling. At $T_{\rm eff} < 10,000$ K, these 
  models predict ages of several Gyr, far exceeding those from SRLOF and 
  minimal-envelope sequences. Flashing sequences significantly extend
  the pre-WD phase compared to non-flashing sequences, but this phase
  remains much shorter than in SRLOF evolution. The amount of $M_{\rm
    H}$ also affects mass and surface gravity estimates, introducing
  systematic differences from SRLOF WDs at a given $T_{\rm eff}$.}
{The evolutionary paths of CE He WDs differ significantly from those
  of SRLOF-produced WDs. Minimal-envelope CE WDs cool rapidly and merge
  at lower temperatures, while those with sustained H burning
  remain bright for longer and merge at higher temperatures. These
  differences with SRLOF WDs are critical for understanding the evolutionary history
  and final fate of He WDs in compact binaries.
  }
\keywords{stars: evolution -- stars: interiors -- stars: white dwarfs -- binaries: close -- stars: low-mass -- methods: numerical} 
  \titlerunning{CE Evolution and Cooling of He WDs}   
  \maketitle


\authorrunning{Althaus et al.}  


\section{Introduction}
\label{intro}  

Helium-core white dwarfs (He WDs) are compact remnants with stellar
masses $\lesssim 0.45\,M_{\odot}$ \citep{2013osp..book.....C}, formed
due to enhanced mass loss before the onset of the core He flash in
low-mass red giant branch (RGB) stars. This mass loss typically
results from binary interactions, although in rare cases, single
evolution can lead to relatively massive He WDs under specific
conditions of low initial mass, low metallicity, and high He content
\citep{2008ApJ...673L..29C,2013ApJ...769L..32B,2017A&A...597A..67A}.

A particularly relevant subclass is the extremely low-mass (ELM) WDs,
with masses below $\sim 0.30\,M_{\odot}$. These objects must form in
binaries, as their masses are insufficient for He ignition
\citep{2013osp..book.....C}. Large observational efforts, including
SPY and WASP surveys, have significantly expanded the catalog of known
ELM WDs
\citep{2009A&A...505..441K,2010ApJ...723.1072B,2012ApJ...744..142B,2011MNRAS.418.1156M,2013ApJ...769...66B,2014ApJ...794...35G,2015ApJ...812..167G,2016ApJ...818..155B,2020ApJ...889...49B}.

The binary origin of ELM WDs is reinforced by the fact that most are
found in compact binaries, predominantly He+CO WD systems
\citep{2020ApJ...889...49B}. Their observed mass-period distribution
suggests two main formation channels: stable Roche-lobe overflow (hereinafter SRLOF) and
common envelope (CE) evolution, triggered by dynamically unstable mass transfer
\citep{2020ApJ...889...49B}. ELM WDs
with masses $ \sim 0.20 - 0.32\,M_{\odot}$ and orbital periods $P <
0.1$ days are likely CE products and expected to merge. Theoretical
models \citep{2019ApJ...871..148L} indicate that the energy required
to eject a tightly bound CE when the donor has just finished core H
burning \citep[see also][]{2018ApJ...858...14S} prevents the formation
of ELM WDs with masses below $0.22\,M_{\odot}$, which instead form via
the SRLOF channel.

The total H content, $M_{\rm H}$, plays a crucial role in determining
cooling evolution of He WDs, particularly those formed through a CE phase \citep{2018A&A...620A.196C,2019A&ARv..27....7C}.
A recent study by \citet{Scherbak2023} also highlighted  the relevance of 
$M_{\rm H}$ for the evolution of post-CE He WDs, showing that different envelope 
masses lead to significant variations in cooling times and structural properties.
Observations suggest that He WDs formed through CE evolution may have 
significantly lower $M_{\rm H}$ than predicted by SRLOF models 
\citep{2009ApJ...699...40S,2021A&A...650A.102I,2019ApJ...871..148L,2018A&A...620A.196C}.
Despite the uncertainties in CE evolution \citep{2016MNRAS.462..362I,2016MNRAS.460.3992N}, 
numerical studies indicate that most of the RGB progenitor’s envelope is lost during CE ejection 
\citep{ivanova2011,sand2020}.

Due to the complexity of CE evolution \citep{ivanova2013review}, most
observational studies determining ELM WD properties rely on SRLOF-based
evolutionary sequences
\citep{2013A&A...557A..19A,2016A&A...595A..35I}, which are widely used
to infer mass and age. Here, the WD
retains a relatively thick H envelope that sustains stable H
burning. However, if CE evolution systematically leads
to lower  $M_{\rm H}$, post-CE He WDs may have cooling properties and
internal structures significantly different from those inferred using
SRLOF models, potentially introducing systematic biases in observed
stellar parameters.

\begin{table}
\caption{Relevant quantities for selected He WD sequences resulting from CE.}
\centering
\setlength{\tabcolsep}{2pt} 
\renewcommand{\arraystretch}{0.9} 
\resizebox{\columnwidth}{!}{%
\begin{tabular}{llll|llll}
\hline
\hline
\\[-0.3cm]
$M_{\rm WD}$ & $M_{\rm H}$ & $\tau_{\rm adjust}$ & $R^{\rm exp}$ & $M_{\rm WD}$ & $M_{\rm H}$ & $\tau_{\rm adjust}$ & $R^{\rm exp}$ \\
$(M_{\sun})$ & $(M_{\sun})$ & (Myr) & $(R_\odot)$ & $(M_{\sun})$ & $(M_{\sun})$ & (Myr) & $(R_\odot)$ \\
\\[-0.3cm]
\hline
\\
0.4352 & $1.27(-3)$ & 0.09 & 1.33  & 0.2724 $\dagger$  & $4.36  (-3)$ & 10.0 & 1.4  \\
0.4352 & $1.15(-3)$ & 0.09 & 0.67  & 0.2724 $\dagger$ & $4.08  (-3)$ & 8.9 & 1.03  \\
0.4352 & $1.04(-3)$ & 0.07 & 0.44  & 0.2724 $\dagger$ & $3.58  (-3)$ & 5.0 & 0.6  \\
0.4352 & $7.81(-4)$ & 0.05 & 0.22  & 0.2724 $\dagger$ & $3.23  (-3)$ & 4.0 & 0.37  \\
0.4352 & $3.81(-4)$ & <$10^{-3}$ & 0.0  &  0.2724 $\dagger$ & $1.39  (-3)$ &  0.01 & 0.0  \\
0.4352 & $3.49(-6)$ & <$10^{-5}$ & 0.0  &  0.2724 & $1.19 (-3)$ & $<0.01$ & 0.0  \\
&                       &            &      &  0.2724 & $1.01 (-3)$ & $<0.01$ & 0.0  \\
&                       &            &      &  0.2724 & $2.0 (-5)$ &  <$10^{-4}$ & 0.0  \\
\\
0.3630 $\dagger$  & $2.14  (-3)$ & 0.51 & 1.37  & 0.2390 $\dagger$ & $6.64  (-3)$ & 70 & 1.0  \\
0.3630 $\dagger$  & $2.04  (-3)$ & 0.48 & 1.09  & 0.2390 $\dagger$  & $5.95 (-3)$ & 63.5 & 0.58  \\
0.3630 $\dagger$  & $1.90  (-3)$ & 0.45 & 0.60  & 0.2390 $\dagger$ & $3.65  (-3)$ & 16.8 & 0.25  \\
0.3630 $\dagger$  & $1.43  (-3)$ & 0.30 & 0.25  & 0.2390  $\dagger$& $2.84  (-3)$ & 15.5 & 0.14  \\
0.3630 $\dagger$  & $8.14 (-4)$ & $<0.01$ & 0.0 & 0.2390 $\dagger$ & $2.11  (-3)$ & 0.1 & 0.0  \\
0.3630  & $7.13 (-4)$ & $<0.01$ & 0.0 &  0.2390  & $1.62  (-3)$ & 0.07 & 0.0  \\
0.3630  & $5.28 (-4)$ & $<0.01$ & 0.0 &  0.2390  & $7.22  (-4)$ & 0.03 & 0.0  \\
0.3630  & $5.02 (-6)$ &  <$10^{-5}$  & 0.0 &  0.2390  & $2.50  (-5)$ &  <$10^{-4}$  & 0.0  \\
\\
0.3208 $\dagger$ & $2.82  (-3)$ & 1.74 & 0.90 &  0.2026 $\dagger$& $8.65  (-3)$ & 230 & 1.65  \\
0.3208 $\dagger$ & $2.02  (-3)$ & 1.04 & 0.32 &   0.2026 $\dagger$ & $7.16  (-3)$ & 198 & 0.81  \\
0.3208 $\dagger$ & $1.75  (-3)$ & 0.82 & 0.23 &   0.2026 $\dagger$ & $6.0  (-3)$ & 160 & 0.4  \\
0.3208 $\dagger$ & $1.26  (-3)$ & $<0.01$ & 0.0 &   0.2026 $\dagger$ & $4.8  (-3)$ & 115 & 0.23  \\
0.3208 $\dagger$ & $1.10  (-3)$ & $<0.01$ & 0.0 & 0.2026 $\dagger$ & $3.25  (-3)$ & 100 & 0.18  \\0.3208 $\dagger$ & $1.04 (-3)$ & $<0.01$ & 0.0 &   0.2026 $\dagger$ & $2.5  (-3)$ & 0.23 & 0.0  \\
0.3208  & $9.09 (-4)$ & $<0.01$ & 0.0 &  0.2026 & $2.06  (-3)$ & 0.18 & 0.0  \\
0.3208  & $6.55 (-4)$ & $<0.01$ & 0.0 &  0.2026 & $1.57  (-3)$ & 0.16 & 0.0  \\
0.3208  & $6.60 (-6)$ &  <$10^{-4}$  & 0.0 &   0.2026 & $7.75  (-4)$ & 0.07 & 0.0  \\
&                     &           &    &       0.2026 & $3.19  (-5)$ &  <$10^{-4}$  & 0.0  \\
\hline
\end{tabular}%
}
\tablefoot{$M_{\rm WD}$: WD mass, $M_{\rm H}$: initial mass of residual H content after progenitor star's envelope ejection.
  $\tau_{\rm adjust}$: adjustment time from CE ejection to maximum \teff. $R^{\rm exp}$: stellar radius after
  envelope expansion due to H ignition. $\dagger$: occurrence of H shell flashes on the cooling branch.
  The negative numbers in parentheses represent powers of 10.}
\label{tabla1}
\end{table}

\begin{figure}
        \centering
        \includegraphics[width=1.\columnwidth]{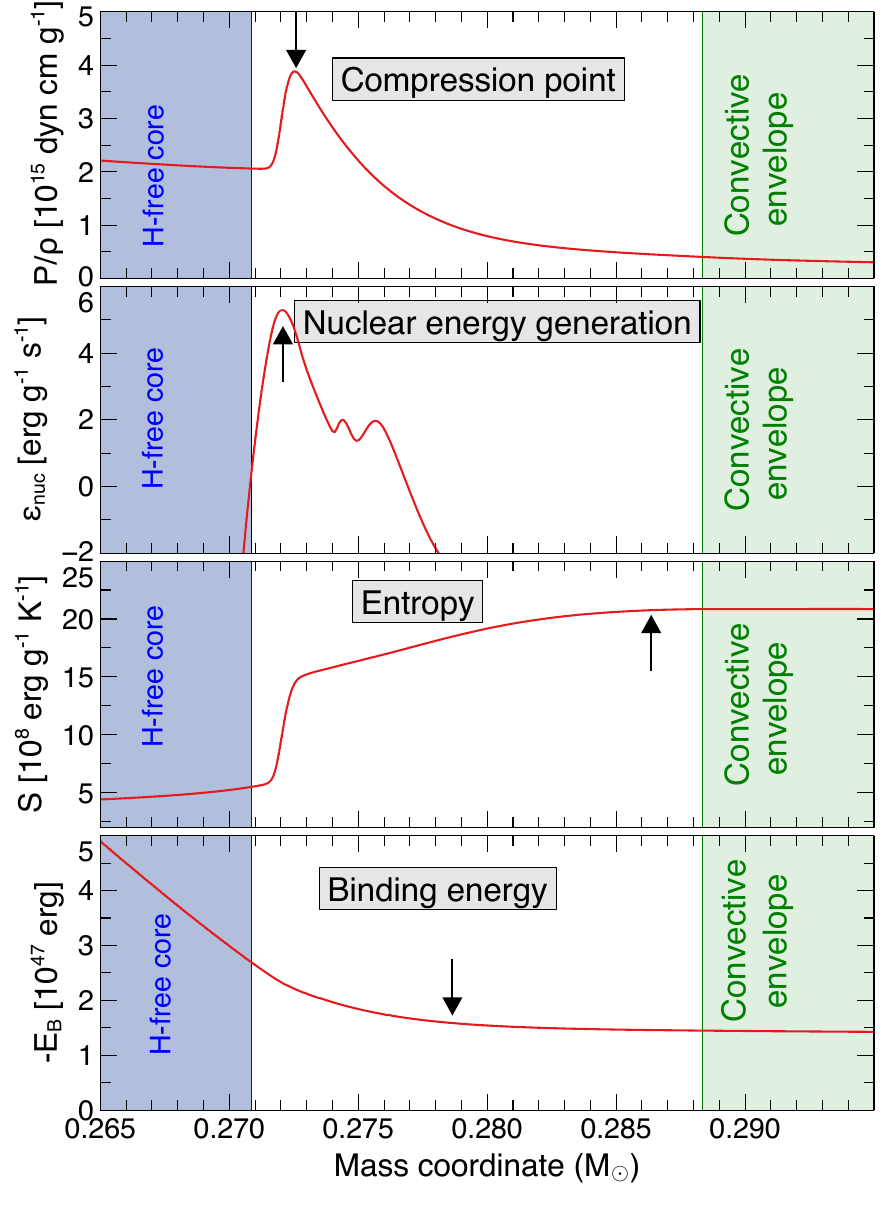}
         \vspace{-0.6cm} 
       \caption{Internal profiles of $P/\rho$, nuclear energy
         generation, specific  entropy, and binding energy for a $1\,M_{\odot}$
         RGB pre-CE star with a H-free core mass of
         $0.271\,M_{\odot}$. Colored areas indicate the convective
         envelope and the H-free core, with the core boundary at $X_H
         = 10^{-6}$. Arrows mark the expected bifurcation point
         separating the remaining core from the ejected envelope.}
        \label{entropy}
\end{figure}

In this work, we present a self-consistent set of post-CE evolutionary 
sequences for He WDs,  covering a mass range of
$0.2 - 0.45\,M_{\odot}$. While \citet{Scherbak2023} also explored the impact 
of envelope mass on WD evolution, our approach differs in both motivation 
and methodology. We start from physically motivated CE scenarios and explore 
distinct mechanisms for envelope ejection—such as dynamical removal near the 
bifurcation point and delayed ejection after envelope expansion—assessing their 
plausibility and impact on $M_{\rm H}$. Our sequences are designed to directly 
link observed $T_{\rm eff}$ and $\log g$ to mass and age estimates for He WDs 
formed via CE, providing a practical tool for cases where standard SRLOF-based 
models may not apply.

Recent works have shown that even within the framework of SRLOF
evolution, the mass–orbital period relation for He WDs can exhibit
substantial dispersion depending on physical processes such as wind
mass loss and the mass-transfer prescription adopted in evolutionary
models \citep{2021MNRAS.502..383Z,2023MNRAS.525.2605G}. In this
context, our CE evolutionary sequences provide an independent avenue
to interpret He WD binaries, especially in the short-period regime
where canonical SRLOF predictions may not apply, and where a clear
mass–period relation might not be expected.

This paper is structured as follows: Sect. 2 describes the input
physics and numerical methods. Sect. 3 presents our results,
including the structural properties and evolutionary timescales of
post-CE He WDs, and Sect. 4 summarizes our conclusions.

\begin{figure}
        \centering
         \includegraphics[width=1.\columnwidth]{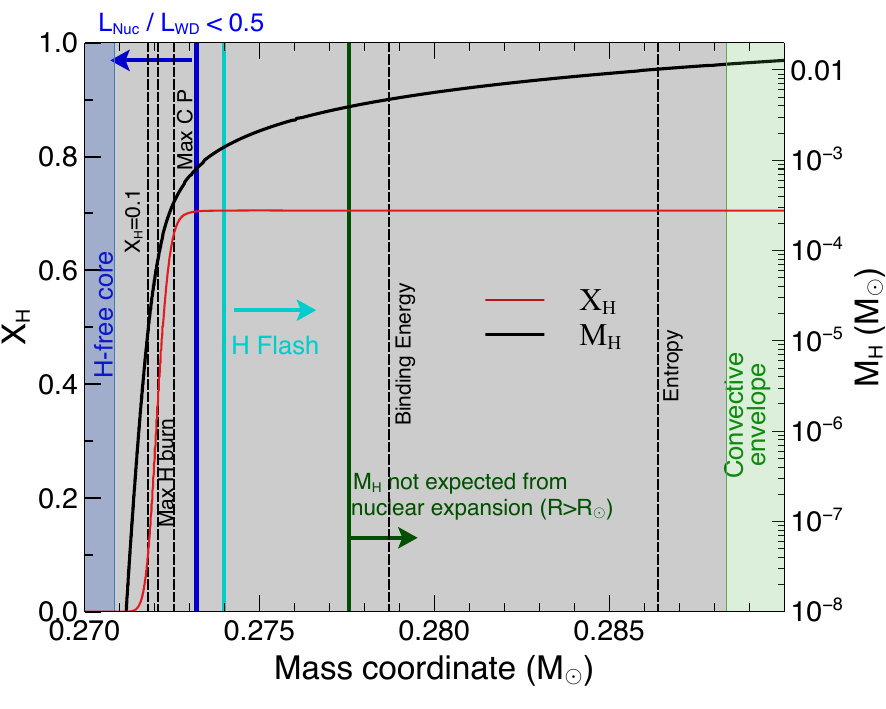}
         \vspace{-0.6cm} 
       \caption{Internal structure of a $1\,M_{\odot}$ pre-CE RGB star with an H-free core 
of $0.271\,M_{\odot}$. The grey region corresponds to the envelope between the H-free 
core and the base of the convective zone, where the bifurcation point separates 
the future WD core from the ejected material. Dashed vertical lines indicate local 
criteria for selecting the bifurcation point: $X_{\rm H} = 0.1$, maximum nuclear 
energy generation rate, maximum compression, and the binding energy and entropy 
conditions (see Fig.~\ref{entropy}). Red and black lines show the H profile and 
its cumulative value $M_{\rm H}$. The dark green line marks the innermost mass 
coordinate for which the residual $M_{\rm H}$ is sufficient to produce envelope 
expansion induced by nuclear burning  beyond $1\,R_{\odot}$. The blue and cyan lines indicate the 
minimum $M_{\rm H}$ required for residual H burning to significantly impact 
cooling, and for the occurrence of unstable H flashes, respectively.}
        \label{bifurcation}
\end{figure}

\section{Creation of stellar models from envelope ejection and their evolution to the WD state}  
\label{formation}

\subsection{Numerical treatment of $M_H$ in post-CE He WDs}

We analyze the evolution of He WD sequences following their formation
in a CE event. To construct them, we evolved a $1\,M_{\odot}$, $Z =
0.01$ star from the zero-age main sequence to the RGB, removing most of the outer envelope at a specified core mass
to simulate the CE phase. The resulting remnant was then evolved from its RGB state to
later He WD cooling stages, considering potential residual nuclear burning.
We used the stellar evolution code {\tt LPCODE}, developed by the La Plata group
\citep{2005A&A...435..631A, 2013A&A...555A..96S, 2015A&A...576A...9A, 2016A&A...588A..25M, 2022A&A...663A.167A}.
Our sequences cover the range $0.20 - 0.45\,M_{\odot}$, consistent with previous studies
suggesting that He WDs formed from CE evolution typically have masses above
$M_{\rm WD} > 0.20\,M_{\odot}$ \citep{2017MNRAS.467.1874C, 2018ApJ...858...14S, 2019ApJ...871..148L}, as
already mentioned. The resulting masses are listed in Table~\ref{tabla1}.

\begin{figure}
        \centering
        \includegraphics[width=1.\columnwidth]{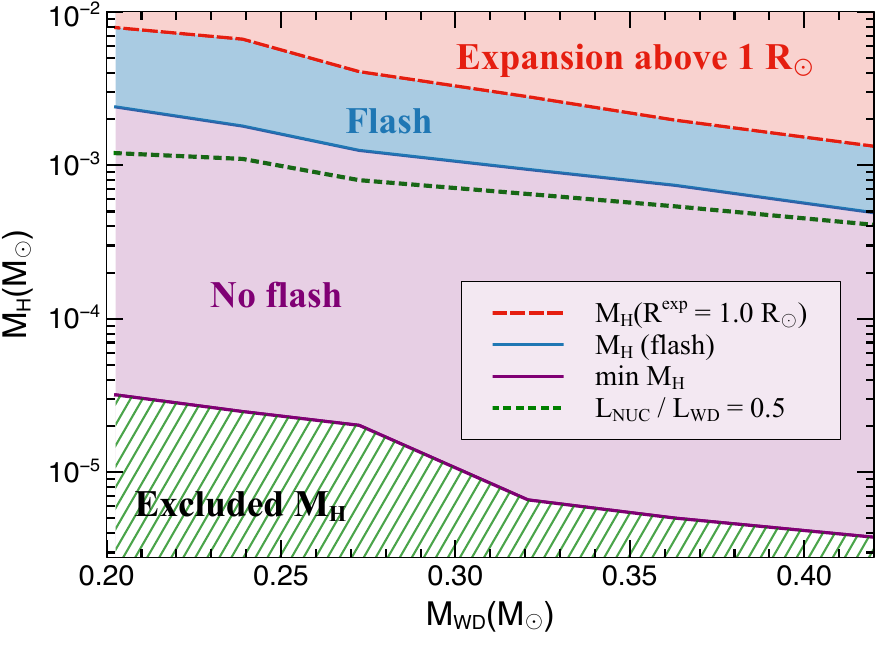}
         \vspace{-0.6cm} 
        \caption{Initial H content mass after CE, $M_H$, as a
          function of WD mass. The dashed red line marks the value of
          $M_{\rm H}$ above which  H burning releases enough
          energy to cause the envelope to expand beyond
          $1\,R_{\odot}$.  Models with larger $M_{\rm H}$ are excluded
          from our sequences. The
           solid blue line separates sequences that undergo or avoid
           H-shell flashes. The green dashed line indicates the $M_H$
           threshold below which residual H burning contributes less
           than $50 \%$ to WD luminosity. The hatched green region
           shows $M_H$ values we excluded from our study.}
        \label{mh}
\end{figure}

\begin{table*}
  \small
\caption{Stellar parameters of selected CE He WDs inferred from post-CE sequences.}
\centering
\setlength{\tabcolsep}{4pt} 
\begin{tabular}{cccccccc|cc}
\hline
\hline
\\[-0.3cm]
Object & $P$  & $T_{\rm eff}$ & log $g$  & $M_{\rm WD}$\, ($M_{\rm H}^{\rm Min}$) &
age($M_{\rm H}^{\rm Min}$) & $M_{\rm WD}$\, ($M_{\rm H}^{\rm Max}$) & age($M_{\rm H}^{\rm Max}$) &   $M_{\rm WD}$(SRLOF) &
age(SRLOF) \\
 & (days) & (K) & (cm s$^{\rm -2}$) & ($M_{\sun}$) & (Myr) & ($M_{\sun}$) & (Myr) &   ($M_{\sun}$) & (Myr) \\
\\[-0.3cm]
\hline
\\
J0651+2844 &  0.00886 & 16340 $\pm$ 260  &  6.810 $\pm$  0.050 &  0.246 $\pm$ 0.008  & 10.47 $\pm$ 1.89   &  0.266 $\pm$ 0.004  & 159.1 $\pm$ 23.5  & 0.251 $\pm$ 0.010  & 188.0 $\pm$ 104 \\
J0822+3048 &  0.02801 & 13921 $\pm$ 160  &  7.146 $\pm$  0.050 &  0.294 $\pm$ 0.011   & 111.5 $\pm$ 18.8  &  0.317 $\pm$ 0.01   & 430.5 $\pm$ 29.6 & 0.311 $\pm$ 0.011 & 233.6 $\pm$ 30 \\
J0825+1152 &  0.05819 & 27180 $\pm$ 400  &  6.600 $\pm$  0.040 &  0.250 $\pm$ 0.005  & 0.412 $\pm$ 0.035  &  0.299 $\pm$ 0.006  & 0.318 $\pm$ 0.04 & 0.287 $\pm$ 0.012 & 83.8  $\pm$ 88 \\
J0923+3028 &  0.04495 & 18761 $\pm$ 220  &  6.860 $\pm$  0.044 &  0.263 $\pm$ 0.007  & 4.22 $\pm$ 0.38    &  0.292 $\pm$ 0.006  & 38.18 $\pm$ 14.3   & 0.278 $\pm$ 0.010  & 72.3  $\pm$ 75 \\
J0935+4411 &  0.01394 & 21660 $\pm$ 380  &  6.960 $\pm$  0.050 &  0.290 $\pm$ 0.011   & 3.37 $\pm$ 1.13   &  0.330 $\pm$ 0.007   & 17.7 $\pm$ 10.7  & 0.318 $\pm$ 0.012 & 2.13 $\pm$ 48 \\
J1053+5200 &  0.04258 & 16370 $\pm$ 240  &  6.540 $\pm$  0.040 &  0.208 $\pm$ 0.006   &  4.87 $\pm$ 0.81  &  0.245 $\pm$ 0.0034   &  37.71 $\pm$ 18.2 & 0.213 $\pm$ 0.007 & 357.6 $\pm$ 151 \\
J1056+6536 &  0.04351 & 21010 $\pm$ 360  &  7.100 $\pm$  0.050 &  0.318 $\pm$ 0.011   & 8.28 $\pm$ 3.45   &  0.346 $\pm$ 0.007   & 53.53 $\pm$ 13.4  & 0.338 $\pm$ 0.012 & 9.2 $\pm$ 36 \\  
J1152+0248 &  0.09952 & 19734 $\pm$ 220  &  7.372 $\pm$  0.043 &  0.376 $\pm$ 0.013   & 52.55 $\pm$ 6.27  &  0.398 $\pm$ 0.012   & 123.11 $\pm$ 5.92 & 0.398 $\pm$ 0.011 & 85.3 $\pm$ 11.6 \\   
J1234-0228 &  0.09188 & 17800 $\pm$ 260  &  6.610 $\pm$  0.040 &  0.224 $\pm$ 0.005  & 3.93 $\pm$ 0.48    &  0.259 $\pm$ 0.004  & 6.92 $\pm$ 2.22   & 0.229 $\pm$ 0.008 & 284.6 $\pm$ 140 \\  
J1235+1543 &  0.03674 & 21024 $\pm$ 310  &  7.178 $\pm$  0.044 &  0.335 $\pm$ 0.008  & 16.6 $\pm$ 5.51     &  0.357 $\pm$ 0.006  & 73.8 $\pm$ 12.0    & 0.354 $\pm$ 0.011 & 21.2 $\pm$ 28.2 \\ 
J1436+5010 &  0.04582 & 17370 $\pm$ 250  &  6.660 $\pm$  0.040 &  0.229 $\pm$ 0.005   & 5.04 $\pm$ 0.61   &  0.261 $\pm$ 0.004   & 15.21 $\pm$ 26.0  & 0.233 $\pm$ 0.008 & 244.7 $\pm$ 140 \\ J1630+4233 &  0.02767 & 16070 $\pm$ 250  &  7.070 $\pm$  0.050 &  0.290 $\pm$ 0.010  & 41.55 $\pm$ 12.7    &  0.306 $\pm$ 0.099  & 228.57 $\pm$ 15.1 & 0.305 $\pm$ 0.011 & 128.4 $\pm$ 20 \\  
J1738+2927 &  0.04770 & 12018 $\pm$ 230  &  6.972 $\pm$  0.051 &  0.249 $\pm$ 0.009  & 134.4 $\pm$ 19.0   &  0.265 $\pm$ 0.008  & 707.99 $\pm$ 75.4  & 0.262 $\pm$ 0.011 & 349.4 $\pm$ 71 \\  
J2338-2052 &  0.07645 & 16620 $\pm$ 280  &  6.850 $\pm$  0.050 &  0.254 $\pm$ 0.008   & 10.16 $\pm$ 1.77  &  0.270 $\pm$ 0.005   & 168.57 $\pm$ 19.5 & 0.262 $\pm$ 0.012 & 150.2 $\pm$ 77 \\  
 \hline
\end{tabular}
\tablefoot{The second column lists the binary's orbital period. The third and
  fourth columns show $T_{\rm eff}$ and $\log g$ of He WDs with $P <
  0.1$ days, likely products of the CE phase
  \citep{2020ApJ...889...49B}. Columns five through eight provide
  stellar mass and age since the CE phase ended, based on non-flashing sequences
  with minimum $M_{\rm H}$ (no residual H burning) and maximum $M_{\rm
    H}$ (enhanced residual H burning). The final
  two columns present stellar mass and age since the end of SRLOF,
  using He WD models from \cite{2013A&A...557A..19A}.}
\label{tablenonuc}
\end{table*}

Theoretical models indicate that after a dynamical CE event, some H
remains bound to the stellar core rather than being fully expelled
\citep{lombardi2006, ivanova2011}. This final $M_{\rm H}$
is crucial for the evolution of the resulting He
WD. An important uncertainty is the bifurcation point, which separates the
retained core from the ejected envelope and remains under debate
\citep{tauridewi2001, ivanova2013review, 2016A&A...596A..58K,
  2022MNRAS.511.2326V, chenreview}. This point is located between the
H-free core and the base of the convective envelope in the pre-CE
star, within the H-rich layers.

Different criteria have been proposed to determine the bifurcation
point, as reviewed by \cite{tauridewi2001} and \cite{ivanova2013review}.
Figure~\ref{entropy} presents the expected
locations of the bifurcation point for a pre-CE RGB model with an
H-free core mass of 0.271 $M_{\odot}$. The arrows indicate the
predictions from each criterion. Specifically,

\begin{itemize}
    \item The bifurcation point can be placed based on the entropy
      profile, identified by the sharp onset of the flat entropy
      region. In stars with a convective envelope, $R \propto
      M^{-1/3}$, so as the envelope is removed, the star
      expands. During CE, this facilitates the loss of outer layers
      until deeper radiative regions are reached. Expansion ceases
      when the entropy profile decreases inward, marking the
      separation between the core and the ejected envelope.
    \item Another criterion uses the binding energy of the envelope. The
      binding energy at a given $M_{\rm i}$ coordinate is defined as:
      \[
      E_B = \int_{M_{\rm i}}^{M_*} \left(-\frac{G m}{r} + u \right) dm,
      \]
      where $M_{\rm i}$ is the mass coordinate where the envelope
      separation occurs, $M_*$ is the total stellar mass at CE onset, $G$
      is the gravitational constant, $u$ is the specific internal energy,
      and $m$ is the mass within the radius coordinate $r$. The
      bifurcation point is located where $E_B$ transitions from a steep
      increase near the core to a more gradual variation outward
      \citep{han1994, tauridewi2001, ivanova2013review}.
    \item Another approach places the bifurcation point at the mass coordinate
      corresponding to the maximum nuclear energy release in the H-burning
      shell.

    \item A more physically motivated criterion places the bifurcation point at
      the maximum compression point, $m_{\rm cp}$, where the ratio
      $P/\varrho$ reaches a local maximum within the H-burning shell prior
      to CE. If the post-CE remnant has a mass below $m_{\rm cp}$, it
      contracts smoothly. Otherwise, it expands on a local thermal
      timescale, leading to further mass loss until stabilizing near $m_{\rm
        cp}$. Since deeper envelope stripping is unlikely, $m_{\rm cp}$
      provides a meaningful limit for the post-CE structure of He WDs
      \citep{ivanova2011}.
\end{itemize}

Fig.~\ref{bifurcation} shows the four bifurcation points for the same
RGB model, along with the mass coordinate where the H mass fraction is
$X_H = 0.1$, a commonly adopted reference. All points lie between the
base of the convective envelope and the H-free core, as indicated by
the gray region. The entropy profile criterion yields the largest core
mass, whereas the maximum compression and nuclear energy criteria
place the bifurcation point closer to the H-free core, which coincide with a
sharp drop in the H distribution profile.

The maximum $M_{\rm H}$ is set by the base of the convective envelope. In
the RGB model of Fig.~\ref{bifurcation}, this corresponds to $0.012
M_{\odot}$, about 30 times larger than the value at the maximum
compression point and thus less likely \citep{ivanova2011}. 
Although this location represents the theoretical upper limit for the bifurcation point, 
such extreme values are not adopted in our evolutionary sequences, as 
observational constraints impose a much lower upper limit on $M_{\rm H}$.
Indeed, newly formed He WDs from CE events typically
have orbital periods $P < 0.15$ day, often as short as $P < 0.05$ day
\citep{2016ApJ...824...46B, Scherbak2023}, corresponding to
separations of 1–1.2 $R_{\odot}$. If $M_H$ exceeds a critical value,
nuclear expansion before the cooling track causes the stellar radius
to surpass this separation, triggering mass loss that reduces $M_H$
and establishes an upper bound. The dark green line in
Fig.~\ref{bifurcation} marks this threshold, which lies below the
bifurcation point predicted by the entropy profile and binding
energy. Thus, we exclude models with large $M_H$ that produce stellar
radii exceeding $\sim 1 R_{\odot}$ (see Table
\ref{tabla1}), as shown by the dashed red line in
Fig.~\ref{mh}. Depending on the WD mass, the upper $M_H$ limit ranges from $1.2 \times 10^{-3}$
to $8 \times 10^{-3} M_{\odot}$, reflecting intrinsic constraints of
our models \citep{Scherbak2023}.

\begin{figure}
        \centering
        \includegraphics[width=1.\columnwidth]{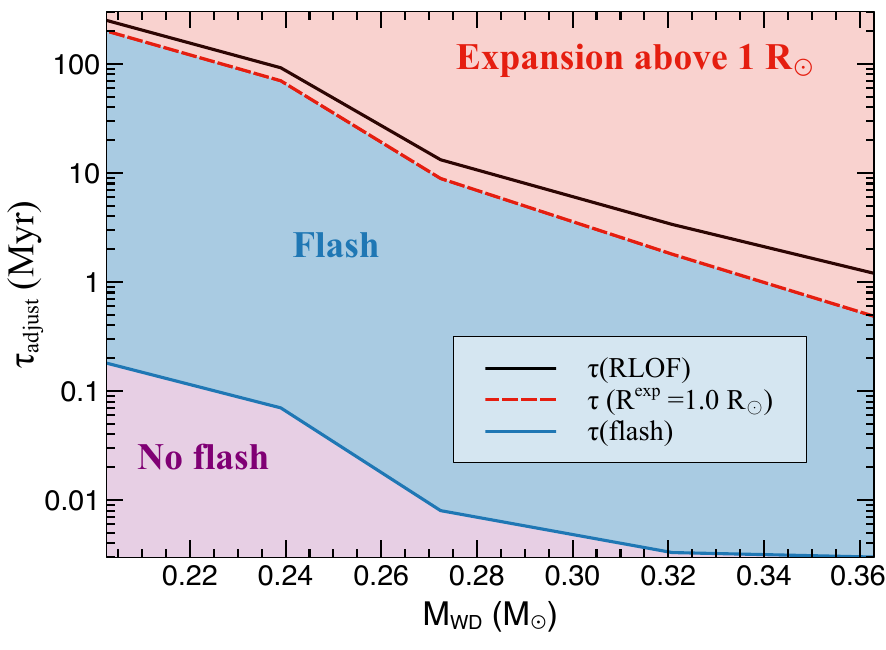}
         \vspace{-0.6cm} 
        \caption{Adjustment time from CE ejection to maximum $T_{\rm
            eff}$ vs. stellar mass. The dashed red line marks
          $\tau_{\rm adjust}$ for sequences where envelope nuclear expansion
          reaches $1\,R_{\odot}$. Non-flashing sequences (below the
          blue line) evolve rapidly to the WD cooling phase. The solid
          black line shows predictions from  SRLOF mass
          transfer.}       
        \label{tau_adjust}
\end{figure}

\begin{figure}
        \centering
        \includegraphics[width=1.\columnwidth]{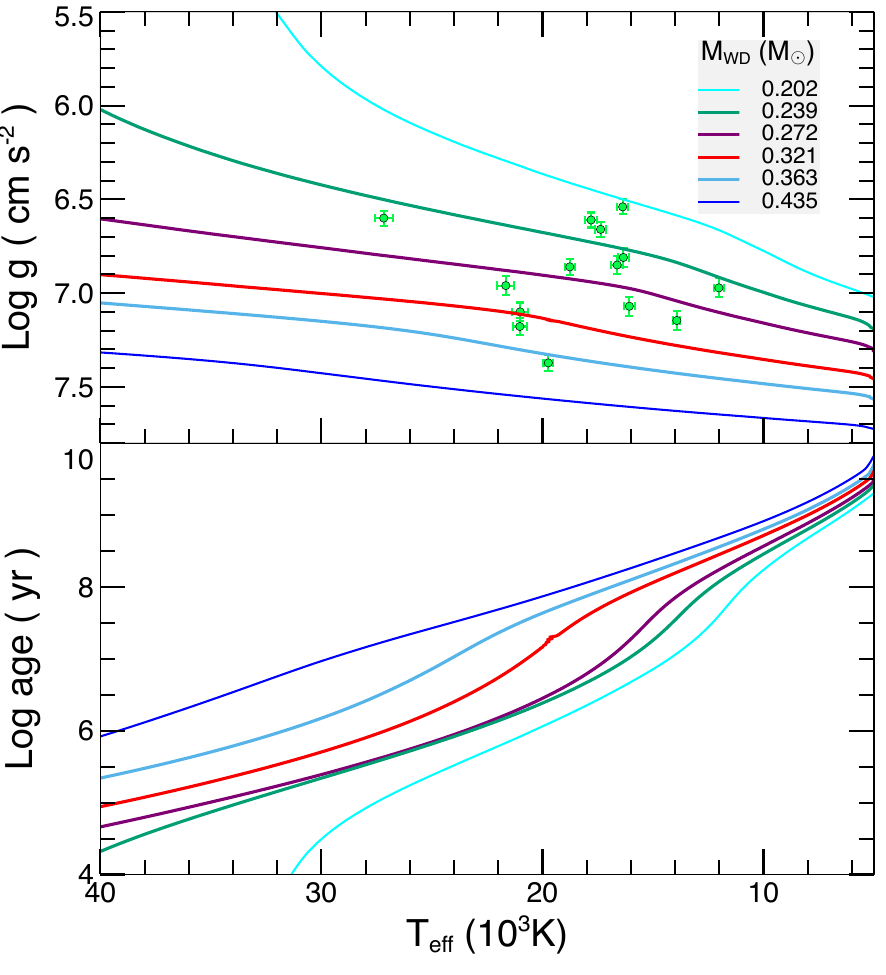}
  \caption{Upper panel: Surface gravity vs. $T_{\rm eff}$ for He WD
  sequences with minimum $M_H$ (no residual H burning). Green symbols
  mark CE He WDs ($P<0.1$ day)  \citep{2020ApJ...889...49B}. Bottom panel:
  Evolutionary time since CE ejection vs. $T_{\rm eff}$. The smooth
  bend at intermediate temperatures results from H diffusion from
  deeper layers, which increases gravity and age.}                     
        \label{gteffnonuc}
\end{figure}

\subsection{The adjustment time to the WD state}
\label{tauadj}

If $M_{\rm H}$ is sufficiently low, the He WD contracts rapidly without 
significant H burning, reaching high $T_{\rm eff}$ in a short adjustment time 
($\tau_{\rm adjust}$) \citep{Scherbak2023}. Unlike WDs formed through 
SRLOF mass loss, these objects enter the cooling track without undergoing 
recurrent H-shell flashes \citep{2013A&A...557A..19A, 2016A&A...595A..35I}, 
defining the so-called “non-flashing sequences.” For moderately larger 
$M_{\rm H}$ values, residual H burning can set in later during the cooling phase, 
providing an additional energy source that slows down the cooling process. 
If $M_{\rm H}$ is increased even further, H burning ignites at earlier stages, 
causing the envelope to expand and eventually leading to the occurrence of
a H-shell flash at the onset of the cooling track, 
which define the “flashing sequences.” Fig.~\ref{mh} (see also Table~\ref{tabla1}) 
shows $M_{\rm H}$ as a function of He WD mass, with the solid blue line marking 
the threshold between flashing and non-flashing regimes.

Fig.~\ref{tau_adjust} shows $\tau_{\rm adjust}$ as a function of He WD
mass, representing the time from CE ejection to the maximum $T_{\rm
  eff}$ before cooling\footnote{In the flashing sequences,
$\tau_{\rm adjust}$ does not include the time spent in the first
cooling track prior to the occurrence of the H flash.}. In all CE sequences, $\tau_{\rm adjust}$ is
markedly shorter than in He WDs from SRLOF mass loss, as
indicated by the solid black line \citep{2013A&A...557A..19A,
  2016A&A...595A..35I}. This is because, in CE sequences, H burning 
is temporarily suppressed due to the rapid mass loss, and resumes only 
after the star has already evolved toward higher $T_{\rm eff}$. 
The delay in H re-ignition shortens $\tau_{\rm adjust}$, especially 
in cases with lower $M_{\rm H}$.

Non-flashing sequences have the shortest $\tau_{\rm adjust}$,
typically below 0.01 Myr, making it negligible compared to cooling
times. In contrast, flashing sequences exhibit longer $\tau_{\rm
  adjust}$, correlated with higher $M_H$ (see Table~\ref{tabla1}), as
early H re-ignition delays entry into the cooling branch. The envelope
expansion limit of 1 $R_{\odot}$ further constrains $\tau_{\rm
  adjust}$ to 0.5–100 Myr, depending on He WD mass. The dashed red
line in Fig.~\ref{tau_adjust} marks this upper bound.

In summary, contraction times to WD state following CE  are
significantly shorter than SRLOF, especially for non-flashing
sequences.

\section{Evolutionary behavior of He WDs formed via common envelope}  
\label{evolutionary}

\subsection{Evolutionary characteristics of He WD sequences with minimal H masses}

The bifurcation point choice strongly affects the final $M_H$ and
subsequent evolution of He WDs. Using the maximum compression point
$m_{\rm cp}$, as proposed by \cite{ivanova2011}, provides a
physically robust criterion to define the post-CE core boundary,
enabling a reliable estimate of $M_H$. As shown in Fig.~\ref{bifurcation}, envelope ejection at $m_{\rm cp}$
yields an $M_H$ where residual H burning contributes less than 50\% of
the total WD luminosity, and is effectively negligible. Thus, He WDs
from CE evolution are expected to rely minimally on residual burning,
with little impact on their cooling times.

While mass removal beyond $m_{\rm cp}$ is unlikely
\citep{ivanova2011}, stellar winds or post-ejection processes may
further reduce $M_H$. Here, we adopt the mass coordinate where $X_H =
0.10$,
following \cite{tauridewi2001,2016A&A...596A..58K}, yielding $M_H$
values about ten times lower than at $m_{\rm cp}$ (see
Fig.~\ref{bifurcation}) without significantly affecting He WD
evolution. The $X_H = 0.10$ criterion enables mass-independent comparisons and
sets a lower $M_H$ limit for our sequences, ranging from $3 \times 10^{-5}$ to $3 \times
10^{-6} M_{\odot}$, depending on He WD mass. Fig.~\ref{mh} highlights
these limits, with the green hatched region marking excluded $M_H$
values. Near-complete envelope removal during CE is supported by 3D
hydrodynamical simulations of low-mass early AGB stars
\citep{sand2020}.

\begin{figure}
        \centering
        \includegraphics[width=1.\columnwidth]{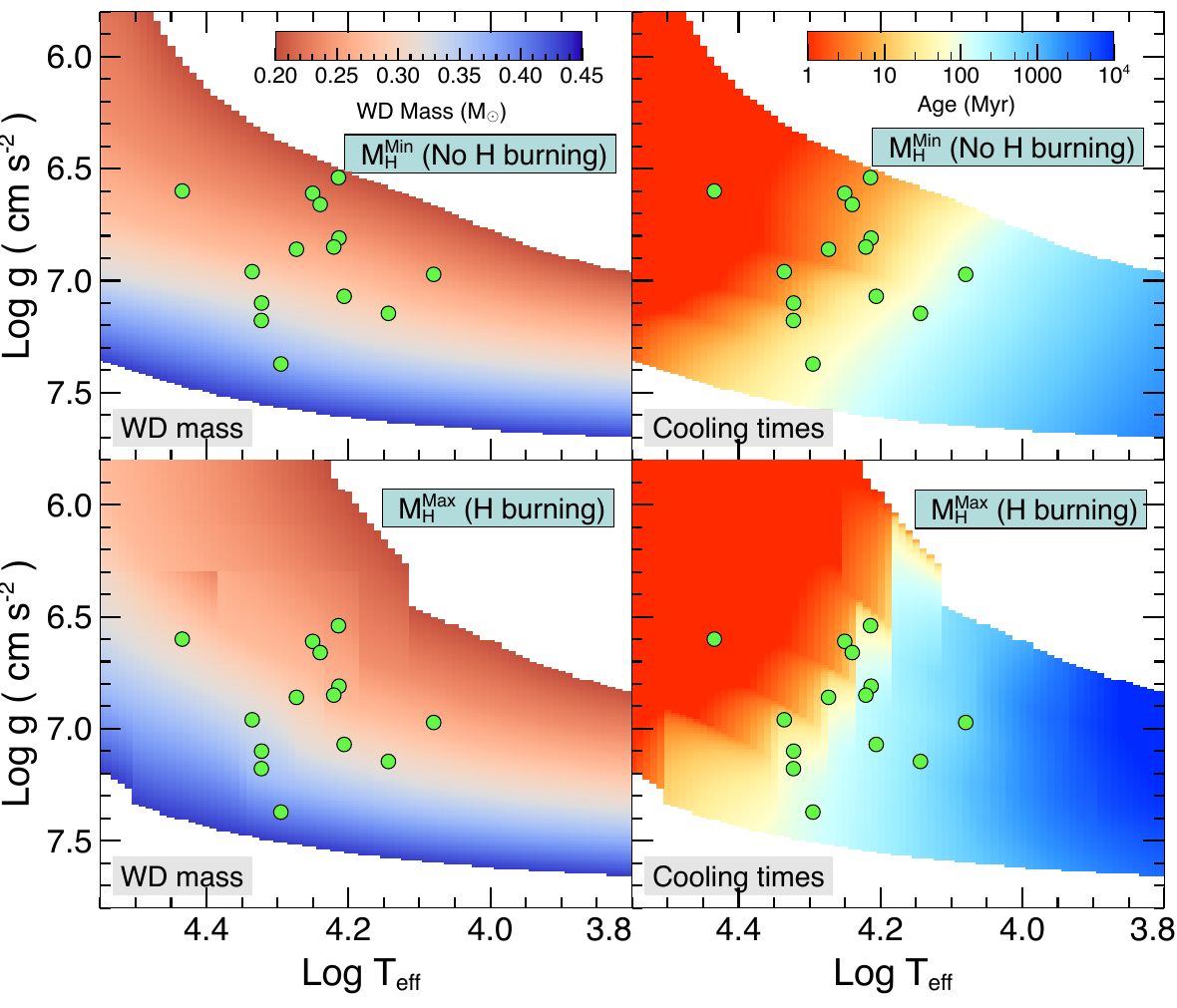}
         \vspace{-0.6cm} 
\caption{Stellar mass (left panels) and cooling times (right panels)
  vs. $\log g$ and $\log T_{\rm eff}$ for post-CE non-flashing
  sequences with minimum $M_H$ (no H burning, top) and maximum $M_H$
  (H burning, bottom). Green symbols mark CE He WDs ($P < 0.1$ day)
  \citep{2020ApJ...889...49B}. Sequences with H burning show much longer
  cooling times, exceeding several Gyr at low $\log T_{\rm eff}$,
  while those with minimum $M_H$ remain below 1 Gyr. In both cases,
  adjustment times after CE remain under 0.1 Myr.}                    
        \label{image}
\end{figure}

The top panel of Fig.~\ref{gteffnonuc} shows surface gravity versus
$T_{\rm eff}$ for He WD sequences with minimum $M_H$. Green symbols
mark He WDs in $P < 0.1$ day binaries from the ELM survey
\citep{2020ApJ...889...49B}, identified as CE products. These objects, listed in
Table~\ref{tablenonuc}, reach the cooling track almost instantaneously
after CE (within a century; see Table~\ref{tabla1}) due to the absence
of nuclear burning. The bottom panel of Fig.~\ref{gteffnonuc} shows
evolutionary times since CE ejection as a function of $T_{\rm eff}$, where cooling is
mainly driven by internal energy depletion in the degenerate He
core. The smooth curve bend at intermediate temperatures results from
H diffusion from deeper layers, which increases surface gravity. As H diffuses upward,
He enrichment at
the envelope's base increases opacity, reducing the energy transfer
rate and prolonging the stellar age.

Sequences with minimum $M_H$ predict ages slightly above 300 Myr for
$T_{\rm eff} < 10,000$ K and several Gyr for $T_{\rm eff} < 5,000$
K. These trends are shown in the upper panels of Fig.~\ref{image},
where stellar mass and evolutionary times are plotted as functions of
$\log g$ and $\log T_{\rm eff}$, alongside observed data. Models
without H burning systematically yield short evolutionary times, even
at low $T_{\rm eff}$, predicting ages below 200 Myr for all observed
CE He WDs, with many under 5 Myr.

\begin{figure*}
        \centering
        \includegraphics[width=1.95\columnwidth]{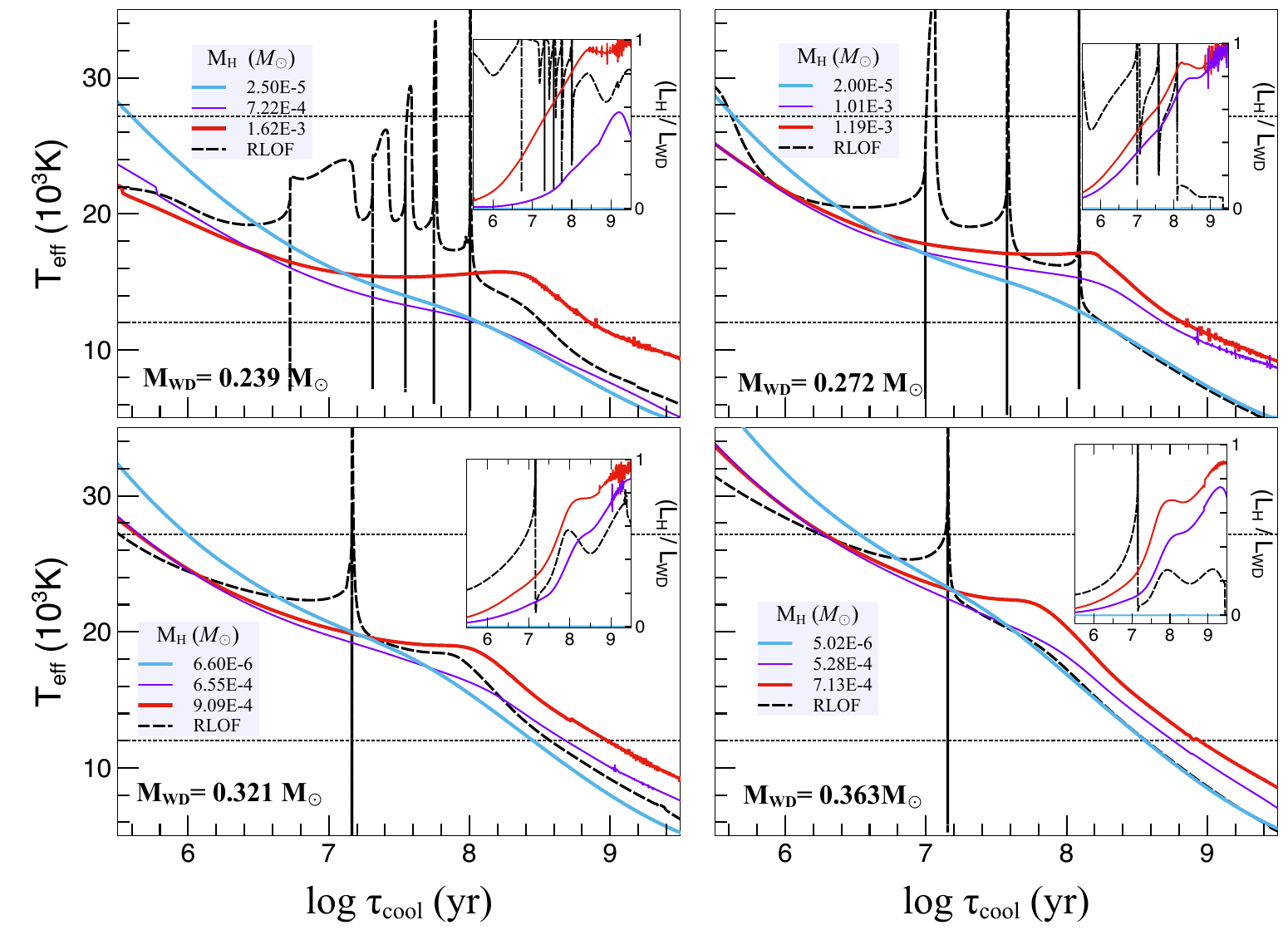}
         \vspace{-0.1cm} 
  \caption{Effective temperature vs. cooling age for non-flashing
  sequences with maximum (red) and minimum (blue) initial $M_H$ from a
  CE phase. Insets show the fraction of WD luminosity from H
  burning. Solid black lines represent SRLOF cooling tracks
  \citep{2013A&A...557A..19A}, where short-lived H flashes occur. Thin violet
  lines correspond to intermediate $M_H$ values. Ages are measured
  from the start of the cooling branch. Horizontal dotted lines mark
  the $T_{\rm eff}$ range of the observed He WDs reported in Table \ref{tablenonuc}.}
        \label{edad_rlof}
\end{figure*}

\begin{figure}
        \centering
        \includegraphics[width=1.\columnwidth]{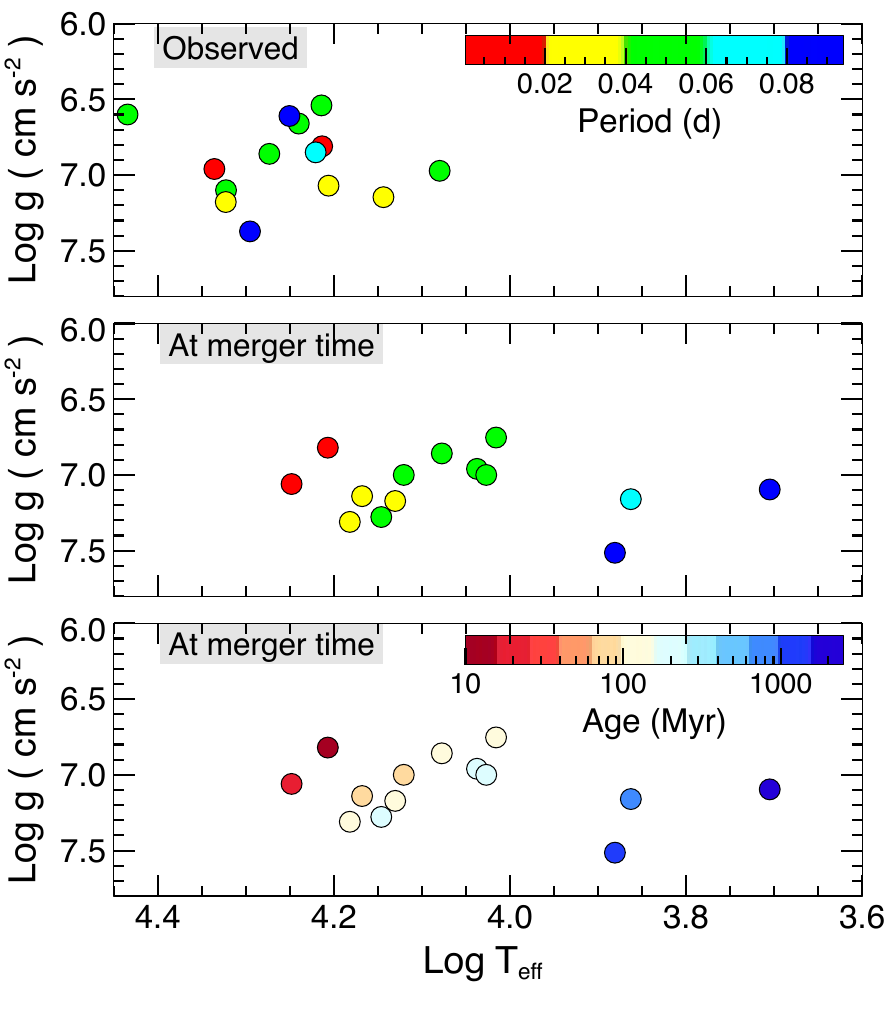}
         \vspace{-0.6cm} 
\caption{Top: Surface gravity vs. $T_{\rm eff}$ for observed He WDs
  with WD companions ($P < 0.1$ day), likely from the CE channel
  \citep{2020ApJ...889...49B}. The color bar shows orbital period. Middle: Same
  systems evolved to merger time, assuming minimum initial $M_H$ (no H
  burning). Bottom: Same as middle, but with the color bar indicating
  WD age at merger. He WDs with $P \gtrsim 0.07$ day are expected to
  merge at $T_{\rm eff} < 8000$ K, reaching $\sim 2$ Gyr.}                        
        \label{gteffmerger}
\end{figure}

\begin{figure}
        \centering
        \includegraphics[width=1.\columnwidth]{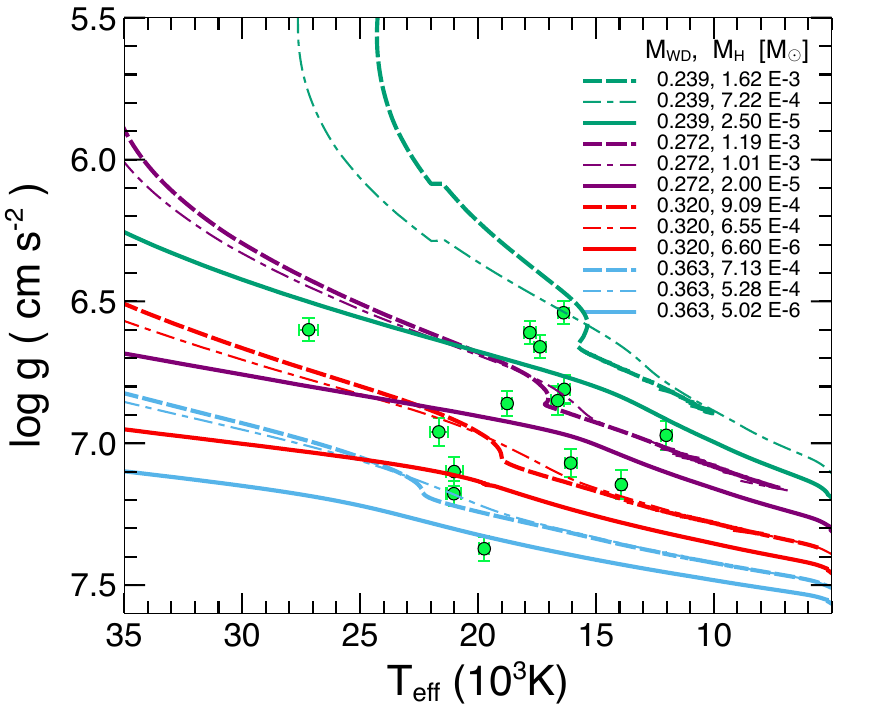}
         \vspace{-0.4cm} 
\caption{Surface gravity vs. $T_{\rm eff}$ for non-flashing He WD
  sequences from a CE phase. Thick solid and dashed lines show minimum
  and maximum $M_H$ sequences, respectively, while thin lines
  represent intermediate $M_H$ values. Green symbols mark CE He WDs
  ($P < 0.1$ day) \citep{2020ApJ...889...49B}.}                       
        \label{gteff}
\end{figure}

Table~\ref{tablenonuc} summarizes the inferred stellar parameters. The
fifth and sixth columns list the stellar mass and  age, both inferred 
from the observed $T_{\rm eff}$ and $\log g$, using sequences with minimum 
$M_{\rm H}$. The ages are measured from the end of the CE phase. These results are 
compared with those derived from SRLOF He WD models 
\citep{2013A&A...557A..19A, 2016A&A...595A..35I}, commonly used to estimate 
WD parameters. The last two columns of Table~\ref{tablenonuc} list the stellar 
mass and cooling age obtained from SRLOF sequences, with ages measured from 
the end of the SRLOF phase. These ages carry large uncertainties due to 
multiple possible solutions arising from early recurrent H flashes 
\citep{2013A&A...557A..19A, 2016A&A...595A..35I}.

For the observed He WDs listed in Table~\ref{tablenonuc}, sequences with minimum 
$M_{\rm H}$ yield slightly lower stellar masses compared to SRLOF models. 
This is because, at a given $T_{\rm eff}$, WDs with thinner H envelopes 
are more compact, and thus require less mass to reproduce the observed surface gravity. 
In addition, CE sequences predict typically younger cooling ages for this sample 
than their SRLOF counterparts, where nuclear burning contributes significantly 
before and during the cooling phase.

To explore the evolution at fixed stellar mass, 
Fig.~\ref{edad_rlof} shows $T_{\rm eff}$ versus cooling age for CE sequences 
with minimum $M_{\rm H}$ (blue lines) and SRLOF sequences (black lines) 
at four different WD masses. For SRLOF models, cooling ages are measured 
from the onset of the first cooling branch. Except for the most massive 
cases, SRLOF models predict  older ages than CE models within 
the $T_{\rm eff}$ range covered by Table~\ref{tablenonuc}. At lower 
temperatures, the age differences decrease, particularly when H flashes 
in SRLOF models reduce the envelope mass enough to quench nuclear burning, 
resulting in cooling times that become more comparable to those of CE WDs.
At earlier stages, near higher $T_{\rm eff}$, CE models with minimum 
$M_{\rm H}$ actually evolve more slowly. This is because their more compact 
structure leads to lower luminosities and slower cooling rates.

The orbits of He+CO WD binaries shrink due to gravitational-wave radiation, with 
the merger timescale determined by the orbital period. We assume that all systems 
in Table~\ref{tablenonuc} will eventually merge \citep{2020ApJ...889...49B}. 
Fig.~\ref{gteffmerger} presents the positions of the observed ELM WDs in the 
$\log g$–$T_{\rm eff}$ plane, both at the current epoch (top panel, based on 
observations) and at the time of merger (middle and bottom panels), as inferred 
from our evolutionary models. The merger positions correspond to sequences with 
minimum $M_{\rm H}$, i.e., no residual H burning. Systems with $P \gtrsim 0.07$ days 
are expected to merge below $8000$ K, at ages approaching $\sim 2$ Gyr (bottom panel). 
Despite their short orbital periods —likely resulting from a CE phase— some He WDs 
may reach low $T_{\rm eff}$ before merging, provided that residual nuclear burning 
is absent.

All sequences with minimal $M_H$ undergo convective mixing at $T_{\rm
  eff} < 5000$ K, leading to H-He envelope enrichment. This occurs in
all He WDs, regardless of mass, as deep mixing transports H from the
envelope into the underlying He layer.

Adopting the maximum compression point $m_{\rm cp}$ as the bifurcation 
point leads to a larger $M_{\rm H}$ than the minimum values discussed 
in this section (see Fig.~\ref{bifurcation}), but does not significantly 
alter the overall evolution of the resulting He WD. At this mass coordinate, 
residual H burning contributes less than 50\% of the total WD luminosity. 
The main difference is due to the fact that the thicker H envelope increases the stellar 
radius and shortens the cooling times by 20–30\% compared to sequences 
with minimum $M_{\rm H}$. This larger envelope also suppresses surface 
convective mixing at low $T_{\rm eff}$.

\subsection{Evolutionary characteristics of He WD sequences with residual H shell burning}

While $m_{\rm cp}$ is a physically motivated bifurcation point, we
examine the potential effects of choosing bifurcation points above $m_{\rm cp}$
on He WD evolution. In massive giants, \cite{2022MNRAS.511.2326V} found
that bifurcation limits above $m_{\rm cp}$ may prevent rapid
re-expansion, potentially setting CE termination thresholds and
yielding higher $M_H$ values. However, since this result applies to
massive stars, we assess its relevance for lower-mass giants without
assuming identical behavior.

Higher $M_{\rm H}$ values significantly impact the evolution of He WDs, 
particularly when residual H burning dominates the star's energy output 
during the cooling phase. In the model shown in Fig.~\ref{bifurcation}, 
this situation arises if the envelope is removed at mass coordinates beyond 
the vertical blue line. In that case, the remaining $M_{\rm H}$ is large 
enough for nuclear burning to contribute more than 50\% of the total 
luminosity once the star reaches the WD stage. 

Fig.~\ref{edad_rlof} (red lines) shows cooling times for models with 
$M_{\rm H}$ values near this threshold —- specifically, the highest values 
that still avoid H flashes, as indicated by the cyan line in 
Fig.~\ref{bifurcation}. Fig.~\ref{edad_rlof}  also shows the fraction of WD luminosity 
powered by residual H burning in these non-flashing sequences. 
This burning prolongs the cooling times significantly below
$T_{\rm eff} \sim 20,000$–$17,000$ K, depending on stellar mass. 
At higher $T_{\rm eff}$, where nuclear burning is negligible, WDs with 
larger $M_{\rm H}$ cool more rapidly due to their higher luminosity and 
larger radii. Conversely, in this hot regime, sequences with minimum 
$M_{\rm H}$ exhibit the longest cooling times at a given stellar mass.

The upper-right and bottom-right panels of Fig.~\ref{image}
illustrate these trends. The sawtooth-like structure in maximum $M_H$
sequences results from H-burning reignition, slowing the cooling rate
and causing nearly constant $T_{\rm eff}$ evolution over a significant
period. Cooling times for maximum $M_H$ sequences can exceed several
Gyr at low $\log T_{\rm eff}$, whereas minimum $M_H$ sequences remain
below 1 Gyr.

The evolution of intermediate $M_H$ sequences depends on whether H
burning remains significant along the cooling track. If sustained,
these sequences yield ages between those of maximum and minimum $M_H$
at both high and low $T_{\rm eff}$ (see Fig.~\ref{edad_rlof}). In summary, at
low $T_{\rm eff}$, maximum $M_H$ sequences evolve the slowest due to
residual H burning, while at high $T_{\rm eff}$, they evolve the
fastest.

As shown in Fig.~\ref{edad_rlof}, sequences with maximum $M_{\rm H}$ predict 
longer cooling times than SRLOF sequences at fixed stellar mass and low to 
intermediate $T_{\rm eff}$, due to sustained H burning. This difference is 
particularly pronounced in the coolest models. A similar effect is often 
seen in the inferred ages of observed WDs in Table~\ref{tablenonuc}, 
though not always. Part of the discrepancy arises because Table~\ref{tablenonuc} 
includes the pre-WD adjustment time—longer in SRLOF models—and the inferred 
masses differ between model sets. As a result, the trends in Fig.~\ref{edad_rlof} 
do not directly translate into those derived from observations.

At higher $T_{\rm eff}$, the behavior changes: both SRLOF and minimum 
$M_{\rm H}$ sequences tend to predict longer cooling times than maximum 
$M_{\rm H}$ models, where the increased radius and luminosity accelerate 
energy loss. This contrast underscores the critical role of H-layer 
thickness and evolutionary history in shaping WD cooling behavior.

The H-layer mass $M_H$ strongly impacts the surface gravity of He WDs,
affecting mass estimates from observed $g$ and $T_{\rm
  eff}$. Fig.~\ref{gteff} illustrates this effect for
non-flashing He WD sequences from CE evolution. Thick and thin lines
represent different $M_H$ values, emphasizing its influence on
evolutionary tracks. Sequences with maximum $M_H$ show a
characteristic "hook" due to H-shell re-ignition, which depletes H and
increases surface gravity as $T_{\rm eff}$ decreases.
Sequences with higher $M_H$ predict larger inferred masses for
observed He WDs compared to those with minimum $M_H$. This trend is
evident in the bottom-left and upper-left panels of
Fig.~\ref{image}, where increased $M_H$ correlates with higher
predicted stellar masses.

In closing this section, we mention that if maximum $M_H$ sequences are considered,
all observed He+CO WD
binaries  \citep{2020ApJ...889...49B} would merge before reaching $T_{\rm eff} =
10,000$ K, preventing them from cooling to lower temperatures.

\subsection{Impact of hydrogen-shell flashes on the evolution of He WDs}

If excess H reignites shell burning before or at the start of the
cooling track, the WD undergoes an early H-shell flash, defining a
flashing sequence. Fig.~\ref{mh} (see also Table~\ref{tabla1}) shows
the $M_{\rm H}$ threshold for this process (solid blue line), which in
Fig.~\ref{bifurcation} corresponds to a bifurcation point above the
cyan line.

If a H flash occurs during the early WD cooling stage, the energy released 
causes the star to expand rapidly, reaching radii larger than $1\,R_{\odot}$. 
This expansion may lead to a renewed episode of SRLOF and 
substantial envelope loss before the star finally settles onto the cooling track, \cite{Scherbak2023}. 
The resulting $M_{\rm H}$ is typically smaller than its pre-flash value and is expected to fall 
between the limits of non-flashing and flashing sequences. From an observational perspective, a WD that 
underwent such a flash episode cannot be distinguished from one that emerged 
directly from the CE with a thinner envelope. However, their evolutionary 
histories differ substantially: flashing sequences experience a significantly 
longer adjustment time $\tau_{\rm adjust}$, while for non-flashing sequences, 
this phase is nearly negligible, as discussed earlier.

For the 0.2026, 0.239, 0.2724, 0.3208, 0.363, and 0.4352 $M_{\sun}$ 
flashing sequences, the maximum pre-WD times from the end of the CE 
are 260, 84, 77, 72, 38, and 0.1 Myr, respectively. Hence, past 
flashes can substantially impact the total age of a He WD. 

Regarding the possibility of detecting a WD just before the H flash—
a phase characterized by slow cooling—the likelihood is low. 
Although this stage lasts relatively long, it occupies a very narrow 
region in the $\log g$–$T_{\rm eff}$ diagram, which limits the chances 
of observing a WD in this evolutionary phase.

\subsection{Impact of progenitor properties on the residual hydrogen mass}

We have also explored the possible impact of progenitor mass and
metallicity on the value of $M_{\rm H}$ resulting from a given
bifurcation point. By evolving stellar models with different initial
conditions until the formation of a He core of fixed mass
($0.27087\,M_\odot$), we found that decreasing the metallicity or
increasing the progenitor mass tends to result in slightly larger
values of $M_{\rm H}$. This effect is moderate when the bifurcation
point is defined by the compression point, which is arguably the most
physically well-motivated criterion. For example, increasing the
progenitor mass from 1 to $2\,M_\odot$ (at $Z = 0.01$) increases
$M_{\rm H}$ at the compression point from $3.5\times10^{-4}$ to
$4.0\times10^{-4}\,M_\odot$, while lowering the metallicity from $Z =
0.01$ to $Z = 0.001$ (for a $1\,M_\odot$ progenitor) leads to $M_{\rm
  H} = 5.0\times10^{-4}\,M_\odot$. In contrast, when using a
bifurcation point defined by a H abundance threshold (e.g., $X_{\rm H}
= 0.1$), the resulting $M_{\rm H}$ becomes significantly more
sensitive to the progenitor structure. In particular, for the
$2\,M_\odot$ case, $M_{\rm H}$ is a factor of 10 larger than in the
$1\,M_\odot$ model, due to the broader H profile shaped by the
convective core during central H burning.

Despite these variations in $M_{\rm H}$, a comparison with our post-CE
sequences shows that the qualitative behavior of He WD evolution
remains unchanged. The differences in $M_{\rm H}$ resulting from
changes in progenitor mass or metallicity primarily affect the cooling
timescales, but not the overall evolutionary path of the WD.
While a detailed study is beyond the scope of this work, these
results suggest that our conclusions are robust against realistic
variations in progenitor properties. Further exploration of these
dependencies may nonetheless help to improve the modeling of the
post-CE envelope structure in future evolutionary calculations.

\section{Conclusions and future work}
\label{conclusions}

This study examines the evolution of He-core white dwarfs (He WDs)
formed via the common envelope (CE) channel, focusing on the influence
of H envelope mass ($M_{\rm H}$). The final $M_{\rm H}$ is determined
by the bifurcation point, marking the boundary between the degenerate
He core and the residual envelope. Selecting the maximum compression
point ($m_{\rm cp}$) results in minimal $M_{\rm H}$, leading to rapid
cooling with negligible residual H burning. If $M_{\rm H}$ exceeds
this minimum, residual H burning becomes significant, extending the
cooling evolution.

We modeled He WDs (0.20--0.42\, $M_{\odot}$) from a 1\, $M_{\odot}$
progenitor in the RGB phase using the LPCODE stellar evolution code,
following their evolution from the CE phase to the cooling track. Our
findings confirm that $M_{\rm H}$ significantly influences He WD
evolution. Consistent with \citet{Scherbak2023}, we identify two
post-CE evolutionary paths:

\textsl{Non-flashing sequences:} These WDs evolve rapidly to the cooling track, 
as their low $M_{\rm H}$ prevents residual H burning. The pre-WD phase in these 
models lasts at most a few centuries. When applied to the observed sample of 
CE He WDs, our sequences predict ages typically below 200 Myr, with many 
younger than 5 Myr. In contrast, WDs formed through stable Roche-lobe overflow (SRLOF) 
experience sustained H burning, resulting in extended cooling times. 

For $M_{\rm H}$ values above the minimum, residual H burning contributes to 
the energy budget, slowing down cooling. When $M_{\rm H}$ is sufficiently large, 
H burning dominates the luminosity for a significant period, especially at 
lower $T_{\rm eff}$. In such cases, our models predict ages exceeding 1 Gyr 
for $T_{\rm eff} < 10,000$K—substantially longer than the ages inferred from 
SRLOF sequences.

\textsl{Flashing sequences:} When $M_{\rm H}$ surpasses a critical
threshold, an H shell flash occurs before the WD reaches the final cooling
track, prolonging the pre-WD phase relative to non-flashing sequences,
yet still shorter than in SRLOF evolution. The flash causes expansion
and additional envelope mass loss, resulting in WDs that settle onto
the cooling track with envelope masses between the minimum and maximum
$M_{\rm H}$ limits of non-flashing sequences.

The value of $M_{\rm H}$ affects the inferred stellar mass.
Sequences with higher $M_{\rm H}$ have larger radii, leading to lower
inferred surface gravities at a given $T_{\rm eff}$ compared to WDs with
minimal $M_{\rm H}$. This effect is crucial for determining the fundamental
parameters of extremely low-mass (ELM) WDs from spectroscopy, as incorrect
assumptions about their evolutionary channel can introduce biases in mass and
age estimates.

We examined the final outcomes of these systems, noting that $M_{\rm H}$ 
strongly influences the $T_{\rm eff}$ at which the merger  occurs.
Binary systems with orbital periods $P \gtrsim 0.07$ days are expected 
to merge at low temperatures ($T_{\rm eff} < 8000$ K) after up to 2 Gyr. 
This occurs in He WDs with minimal $M_{\rm H}$, which cool efficiently 
in the absence of residual nuclear burning. Conversely, He WDs with 
higher $M_{\rm H}$ generally sustain prolonged H burning, leading to mergers at 
higher temperatures and different WD ages at merger. Understanding this distinction is 
essential for linking merger remnants, such as hot subdwarfs and extreme 
He stars, to the broader population of compact binaries.

An important aspect of this work is the provision of tables that enable
direct inference of mass and age for observed CE ELM WDs based on their surface gravity
and $T_{\rm eff}$. These tables offer a reliable reference
for CE-produced WDs, avoiding biases that may arise
when applying SRLOF sequences. Although we
primarily focus on non-flashing sequences, the tables are also appropriate
for He WD flashing sequences; in these cases, it is essential to
consider pre-WD ages.

We also examined how realistic variations in progenitor mass and
  metallicity affect the residual H-envelope mass associated with
  different structural bifurcation points. Although the resulting
  $M_{\rm H}$ values can differ, especially when defined by the
  H-profile, the overall post-CE evolution remains qualitatively
  unchanged across the explored range. These findings reinforce the
  robustness of our conclusions. Note that this is a structural
  analysis and does not account for the energetics required to reach
  such bifurcation points during CE ejection.

Future work should also focus on incorporating rotational dynamics
into the modeling of ELM WDs formed through CE evolution. Tidal
interactions during the CE phase could impart significant angular
momentum to the resulting WDs, potentially leading to rapid
rotation. Rapid rotation can significantly influence internal
processes such as element diffusion, convective mixing, and magnetic
field generation, thereby affecting the cooling rates and pulsation
properties of ELM WDs. Additionally, existing studies have explored
the effects of rotational mixing in low-mass WDs resulting from SRLOF,
highlighting the importance of incorporating rotational effects into
stellar evolution models \citep{2016A&A...595A..35I}. Integrating
rotational dynamics into these models and utilizing asteroseismology
to probe the internal structures of ELM WDs will refine our
understanding of their fundamental properties and evolutionary
pathways.

\begin{acknowledgements}

 We thank the referee, Dr. Xianfei Zhang, for insightful comments that
 improved the quality of the manuscript. We thank the Asociaci\'on
 Argentina de Astronom\'ia for supporting the publication costs of
 this article. This research has made use of NASA Astrophysics Data
 System.
  
\end{acknowledgements}

\bibliographystyle{aa}
\bibliography{biblio}

\end{document}